# Two Major Paths of Gene-Duplicates Evolution

Dmitri Parkhomchuk[§], Sergei Rodin
Department of Theoretical Biology, Beckman Research Institute of the City of Hope, 1500 East Duarte Road, Duarte, CA 91010, USA

[§]Corresponding author

Email addresses:
    DP: DParkhomchuk@coh.org
    SR: SRodin@coh.org

## Summary

**Evolving genomes increase a number of their genes by gene duplications. To escape degradation in a functionless pseudogene, any gene duplicate needs to be guarded by negative (purifying) selection from otherwise inevitable fixation of degenerative mutations. In the present study we focus on the evolutionary stage at which new duplicates come under such surveillance.**

**Our analyses of several genomes indicate that in about 10% gene pairs, selection begins to guard a new gene copy very soon after a duplication event whereas the vast majority (90%) of extra genes remain redundant and unrecognised by selection. Such duplicates accumulate all mutations (including degenerative) in neutral fashion and are actually destined to become pseudogenes. We revealed this "two-stream" evolutionary pattern by the analysis of mutations in $2^{nd}$ versus $3^{rd}$ codon positions but not by the routinely used ratio of amino acid replacements (R) versus silent substitutions (S), i.e. the '$2^{nd}$ vs. $3^{rd}$' metric proved to be more resolving than the traditional 'R vs. S' one for distinguishing neutrally evolving future pseudogenes from their functional counterparts controlled by negative selection.**

**In gene databases for large genomes, hundreds of future pseudogenes are annotated as functional genes because they do look like intact and valuable by standard criteria, including even active transcription and translation.
Apparently, these "pseudogenes-to-be" over-cloud and mimic those very infrequent gene duplicates with increased sequence evolution rates driven by positive selection.**

**Abbreviations:** AA, aminoacid; R, aminoacid replacement substitutions; S, aminoacid silent substitutions; sps, substitutions per site; G-value, number of genes in a genome.

## Background

It was observed (Lynch and Conery 2000) that on the average, new duplicates evolve under different constraints on their sequence than do the old ones – the number of aminoacid (AA) replacements is closer to the number of silent substitutions for new duplicates. Thus it was proposed (ibid.) that new duplicates are under "relaxed constraints", which gradually increase to the "normal" level because an extra gene copy apparently can be redundant after the duplication but recognized by purifying selection later. Another general trend is that the number of new duplicates is much larger than that of more diverged ones (ibid.), implying that the majority of duplicates eventually turn into pseudogenes and perish. In this work we address the following questions: To what extent the "relaxed selection" is actually relaxed and for how long? At what stage does purifying selection recognize long-persisting genes? In order to clarify these aspects of collective behaviour of duplicates *en mass* we analysed several eukaryotic genomes and proposed a model closely describing the observed trends and answering the above questions.

## Results

The pairs of gene duplicates have been collected for analyses as described in Materials and Methods. To evaluate the selection pressure, we used two pair-wise metrics – the number of AA replacements versus the number of synonymous mutations (R vs. S) and the numbers of substitutions in $2^{nd}$ codon positions versus that in $3^{rd}$ ones ($2^{nd}$ vs. $3^{rd}$).

The effect of selection on the base substitutions that have been accumulated in human duplicate gene pairs is shown for both metrics (Fig. 1).
The plot of $2^{nd}$ vs. $3^{rd}$ position seems to be more informative in comparison with the R vs. S plot. One can see that the young duplicates form a group with the significantly steeper slope than that of the older duplicates. The contrast between young and old duplicates is much weaker on the R. vs. S plot. Further we focus mainly on this, more resolving $2^{nd}$ vs. $3^{rd}$ metric, discussing the reasons of the R vs. S inferiority later.

Apparently, the numbers of accumulated mutation in a genes pairs have inherent statistical variance around some mean values, which we are interested in. To reveal the trends of the mean selective constraints we averaged the data of Fig. 1 over the duplicates pairs (Fig. 2). We observed two asymptotes of the averaged selection constraints: one is for young duplicates near the coordinate origin with the slope ~0.8 and the other for old ones with ~0.23 slope (Fig. 2B). We observed the same slopes for all large genomes studied thus suggesting that they reflect some universal constants. We proposed that the steeper slope of young duplicates corresponds to the neutral sequence evolution and the more sluggish slope of older duplicates represents the constraints of purifying selection acting on functional genes. The duplicates throughout their evolution do not switch from one slope to another but follow only one of them from the start. Thus our model is as follows:

The divergence pattern (Figs. 1, 3) suggests two classes of duplicates pairs, one class subject to purifying selection and the other class subject to neutral drift (Fig. 3). In the second (largest) class of pairs, at least one gene in a pair is a future pseudogene ("pseudogene-to-be") and accumulates mutations freely until it is recognized as a pseudogene; at this time, the duplicate is no longer in the duplicates database and disappears from consideration. This class is large for young duplicates but exponentially decays in time (measured in substitutions per site) simply because detectable pseudogenes are excluded from considerations.

The large difference between these two metrics (Figs. 1-3) results from the genetic code: the majority of substitutions in the $2^{nd}$ codon position lead to radical AA replacements, which are accepted in evolution very slowly in comparison with substitutions in the $3^{rd}$ position (of which 70% are synonymous). In contrast, among all base substitutions leading to AA replacements (R) there are many near-neutral changes (like many substitutions to AAs from the same column of genetic code), which are therefore far more acceptable than substitutions in the 2nd position. As a consequence, for the R vs. S metric the average slope of pairs under purifying selection is steeper and closer to the slope representing neutral evolution of the "pseudogene-to-be" duplicates (Figs. 2B, 3). These "neutral" slopes are about the same for both metrics. Below is the mathematical formalism for this model described in terms of 2nd vs. 3rd position divergence.

The density of decaying pseudogenes-to-be depends on the duplicates "age" as measured in substitutions per site (sps):

$$N(s) = N_0 e^{-ks},$$

$N_0$ is the rate of production of pseudogenes-to-be per time unit measured in sps, $s$ - the number of sps and $k$ - decay coefficient.

The pseudogenes-to-be group consists of predominantly (gene, pseudogene-to-be) pairs that drift along the line $y = k_p x$, $k_p \approx 0.8$ (Fig. 3). Notably, $k_p < 1$ because the distance between 3rd codon positions increases faster than that between 2nd positions for (gene, pseudogene-to-be) pair. The commonly used statement "for a pseudogene R/S~1" (or 2nd/3rd~1) is not rigorous, because these metrics are pair-wise so that we have to specify the other member of a pair. For a pseudogene-pseudogene pair 2nd/3rd~1 obviously. However, the "neutral" pairs, which we considered here, consist of a gene and its pseudogene-to-be paralog. The average ratio 2nd/3rd for such pair is less than unity because both of the genes in a pair diverge rapidly at 3rd codon positions but only pseudogene-to-be diverge rapidly in 2nd position.

The gene-gene pairs drift along the line $y = k_{23} x$, $k_{23} \approx 0.23$ (Fig. 2). Both of these slopes (neutral and constrained) are not the arbitrary parameters of our model. Rather they reflect the universal properties of the genetic code and average constraints on protein sequence evolution, which are invariant across species. Indeed, we found a remarkable agreement of these values fitted by our model (Figs. 2-4) with independent measurements. The value $k_p$ can be estimated from the fact that the average substitution rates in the 3rd and 2nd positions of functional genes are about $r_3 = 40\%$ and $r_2 = 10\%$ of the neutral rate, respectively (Ophir et al. 1999). Because both of these sites in pseudogenes-to-be mutate at $r_p = 100\%$ of the neutral rate, the pair (gene, pseudogene-to-be) will have a ratio of 2nd position rate to 3rd (without saturation effects): $k_p = \dfrac{r_2 + r_p}{r_3 + r_p} = \dfrac{0.1 + 1}{0.4 + 1} \approx 0.79$.

The ratio of fixation rates in the 2nd and 3rd codon positions for gene-gene pair is also estimated consistently by the above relative rates:

$k_{23} = \dfrac{r_2 + r_2}{r_3 + r_3} = \dfrac{0.1 + 0.1}{0.4 + 0.4} = 0.25$. Both of these values are perfectly consistent with the two-stream model for the neutral and constrained evolution of gene pairs (Fig. 3).

Under the assumption of continuous and constant duplication production, the total gene density is the sum of the decaying stream of pseudogenes-to-be and the steady stream of functional duplicates (Fig. 3). On the basis of constant density of old duplicates (Fig. 4B) we assume that once fixed, functional duplicates persist forever or are lost negligibly slowly.

The sum of these two streams flow rates represents the duplicates density as a function of sps in the 3$^{rd}$ codon position (Fig. 4B):

$$N(s_3) = N_f (\theta \exp(-ks_3) + 1),\qquad \textbf{Eq.1}$$

$s_3$ - number of sps in 3$^{rd}$ codon positions.

$\theta = \dfrac{N_p}{N_f}$ - ratio of production rates of pseudogenes-to-be to functional duplicates.

The number of sps in 2$^{nd}$ codon position for the pseudogenes-to-be stream is $k_p s_3$ and for functional genes $k_{23} s_3$. Hence their weighted average is (Fig. 4A):

$$s_2(s_3) = s_3 \frac{k_p \theta \exp(-ks_3) + k_{23}}{\theta \exp(-ks_3) + 1}.\qquad \textbf{Eq.2}$$

Fig. 4 and Supplementary Fig. 1 show that these equations closely describe the dynamics of duplicates evolution. Fitting parameters are species-specific gene numbers in two streams and the pseudogenization rate $k$.

To plot the analytical density (Fig. 3), we used a binomial distribution to describe the variance of mutation accumulation. For the functional stream, this variance, besides statistical fluctuations, depends on various protein constrains, since there is some diversity in protein sequence conservation, thus widening the functional stream.

We can estimate a lower theoretical limit of $k$ for pseudogenization by base substitutions, which corresponds to the upper limit of the lifetime of pseudogenes-to-be as they decay into pseudogenes. For this we simulated random mutations in all known human genes and detected the time of appearance of the first stop codon in coding sequences. At this point, the gene turns into a pseudogene and is removed from the set. The first order coefficient of the exponential decrease of intact genes turned out to be $k = 15$. For an average gene of 1 kb this corresponds to half-life of about 40 substitutions – a rather large number and a long time before a pseudogene-to-be becomes identifiable as a pseudogene.

The pseudogenization coefficient $k$ can be easily increased above the theoretically expected minimum value by species-specific insertions and deletions (indel) because the corresponding frameshifts are likely to produce an immediately detectable pseudogene. In mammals, indels happen with a low frequency of 5% of the neutral substitution rate (Cooper et al. 2004), and accordingly they have low $k \approx 18$ close to the theoretical lower limit (~15).

Notably, the comparisons of trends for 1$^{st}$ versus 2$^{nd}$ codon positions and 3$^{rd}$ versus 1$^{st}$ revealed the same patterns with corresponding relative rates coefficients (data not shown). Thus, as expected, the pseudogenes-to-be evolve with equal rates in all 3 codon positions.

On the density plot of R vs. S metric (Fig. 3), the angle between the two streams is so small that they are hardly distinguishable at all: the functional stream starts out at ~0.6 slope and the stream of pseudogenes-to-be at ~0.8 slope (about the same as for 2$^{nd}$ vs. 3$^{rd}$ codon positions). Since the variances in the number of mutations (stream width) make them strongly overlapping, the two-stream pattern is very obscured (Fig. 3). In comparison, the slopes of 2$^{nd}$ vs. 3$^{rd}$ codon positions are ~0.8 and ~0.25. These slopes are demonstrated by asymptotes on Fig. 2. As a consequence, the "bump" caused by young pseudogenes-to-be for the R vs. S metric is much less pronounced than that in the 2$^{nd}$ vs. 3$^{rd}$ metric (Fig. 2).

## Discussion

Our results strongly support the hypothesis that the majority of young duplicates belong to the neutral pseudogenes-to-be group. The following general arguments also evidence in a favour of this hypothesis: i) In the case of full-length gene duplication we are unable to recognize an extra copy as a pseudogene until some apparent defects in sequence appear; ii) The exponential elimination rate of the 90% of new duplicates (Fig. 4B) is consistent with the expected pseudogenization rate by neutral substitutions and indels; iii) The observed average mutations rates in all 3 codon positions for young duplicates are consistent with the neutral fixation of substitutions in (gene, pseudogene-to-be) pairs.

One more evidence is the characteristic saddle region in the dynamics of selection pressure at $2^{nd}$ vs. $3^{rd}$ codon positions (Fig. 2), it strongly suggests that duplicates are not a single homogeneous group with gradually changing selection constraints applied to all of the duplicates. If the duplicates evolved with the slopes consequentially changing from the neutral to constrained one, the negative slope and positive curvature (concave up ∪) of the saddle region could not be observed and the asymptote for the old duplicates could not pass through zero. A trajectory with gradual changes from the weak constraints to the strong ones has positive slope and negative curvature (concave down ∩) everywhere (the signs of the first and second derivatives correspondingly). Averaging over a set of such trajectories could only yield a mean trajectory with the positive slope and negative curvature everywhere (Fig. 5). However in all large genomes studied the saddle region is present invariably. Introducing not sequential but concurrent streams of neutrally evolving decaying group and slowly evolving persisting group solves this paradox naturally enough. Again, the R vs. S metric is so unresolving that it actually permits the interpolation by the concave down curve (Fig. 2A), because its statistical variance is larger than its "saddle region" feature.

The second group of functional duplicates overlaps on divergence plots (Fig. 3) near the coordinates origin with the larger pseudogenes-to-be group due to their statistical variances. At the time the later group is sufficiently eliminated by pseudogenization ($sps$ $3^{rd}$ > 0.2, Figs. 2, 4), the evolution of functional genes is approximated by a straight line (Fig. 2, 3) pointing to coordinates origin. Its asymptotic behaviour implies that it is a straight line from the very beginning (Fig. 2B). Thus the purifying selection recognizes the functional genes right after the duplication events.

For some species, the assumption of constant influx of new duplicates does not hold. For example, *Arabidopsis* experienced a relatively recent polyploidization event or a series of large block duplications (Cannon et al. 2004), and therefore it cannot be described properly with this assumption. By contrast, yeast polyploidization (Kellis et al. 2004) took place too long ago to be observed in these plots.

The differences in $k$ values (Table 1) reflect the differences of pseudogenization rates in species due to different indel rates and probably the increased selection pressure on genome size for small genomes. Remarkably, *Drosophyla* has a very high $k$ value and is known to have an unusually high deletion rate (Petrov et al. 2000), correspondingly a fly genome is compact and has small numbers of both pseudogenes-to-be and already established pseudogenes (Petrov et al. 2000). While yeast also has a quite large $k$, it is probably also due to selection acting for minimization of the genome size (and maximization of replication rate).

The pseudogenization coefficient $k$ being calculated only from a set of predicted genes can be used to estimate the indels rates without detecting the indels themselves. Besides confirmation of the high indel rate in *Drosophyla* from our data, we can predict that puffer fish has a higher indel rate than that of studied mammals. Notably, the lowest $k$ of mammals is near the lower theoretical limit of 15 we determined by simulations. If it has been lower, we could challenge the view that these duplicates are pseudogenes-to-be. However they decay in just the time required (or faster due to pseudogenization by indels) to become detectable pseudogenes.

We can roughly predict the number of pseudogenes in a genome, analysing only the predicted genes sequences (few percents of the large genome total sequence), without the exhaustive search for pseudogenes *per se*: If a genome has a low deletion rate, then the decayed "former pseudogene-to-be" can be found as established pseudogenes for a long time – up to the limit of homology detectability, which is about 1 *sps* (corresponds to about 40% nucleotide identity between neutrally diverged sequences). According to Table 1, for a time of accumulation of 1 *sps* for human, about 22,000 pseudogenes-to-be will appear and turn into pseudogenes, which is rather close to direct estimates (~20,000) of the number of pseudogenes in the human genome (Harrison et al. 2002).

Species with smaller genomes have a smaller fraction ($\theta$) of new pseudogenes-to-be (Table 1). We suggest that in species with smaller genomes, a new pseudogene-to-be is more likely to be disadvantageous to the organism at the time of duplication due to disruption of the dense packing of functional elements in such genomes and possibly a pressure on replication time. Hence, in order to be fixed a new duplicate must bear some positive function from the start and the fixation of pseudogenes-to-be by random drift is reduced in comparison with functional genes in small genomes. Also the lifetime of such pseudogenes is shorter. On the contrary, in genomes with large amounts of "junk" DNA, new insertions are likely to be neutral and fixed by drift, leading to a large portion of pseudogenes-to-be and pseudogenes.

The rate of functional duplicates accumulation (Table 1) looks rather low on the grand evolutionary scale. For example, on the basis of constant rate of fixation of functional duplicates (flat region on Fig. 4B) we can speculate that the rate of functional duplicates fixation was approximately constant over geological times. Then we can interpolate gene number (G-value) back in time from the modern values. Interestingly, it turns out that the time required for large genomes to grow to the modern G-values is roughly 1 billion years (Fig. 6), consistently with the other estimates on the origin of the first complex organisms (Blair Hedges and Kumar 2003). Hypothetically half a billion years ago, our common ancestor with fish had approximately half of the current G-value, but both our lineages eventually increased G-values to the present-day values.

## Conclusions

Our model (Fig. 3) implies that functional duplicates are surveyed soon after the duplication event by selection, consistently with some theoretical models and genome-wide analyses (Gu et al. 2002; Rodin and Riggs 2003; Huminiecki and Wolfe 2004; Rodin and Parkhomchuk 2004). These persisting duplicates accumulate further mutations in the same way as typical genes evolving under constraints of purifying selection. Otherwise duplicates become pseudogenes-to-be that passively drift toward a complete degradation into true pseudogenes and further into junk DNA. Fig. 3 shows this bifurcation as an early split of duplicates pairs in two streams. It seems very unlikely for neutrally mutating presumptive pseudogenes to be captured by selection later, after the fixation of random mutations, as the majority of occurring random mutations are deleterious for protein function. It was found that about 35% of occurring AA substitutions inactivate a protein function (Guo et al. 2004). Even a higher percentage might have milder deleterious effects.

The probability that one in the pair of young duplicates is a pseudogene-to-be is more than 90% for human and other large genomes. This raises an important task of reliable statistical separation of functional genes from their abundant "pseudogenes-to-be" paralogs in large genomes. Probably, some of them can be classified by genetic expression analyses, such as that conducted in (Mounsey et al. 2002) suggesting that young duplicates representing about one fifth of nematode annotated genes are actually pseudogenes (mostly pseudogenes-to-be in our terms). However, preliminary patterns for detectably expressed duplicates (from cDNA databases) appeared to be virtually the same as those for predicted genes (Fig. 2B). Apparently, the transcription machinery also does not distinguish valuable genes from pseudogenes-to-be and transcribes the latter until their sequences for transcription initiation degrade or they get silenced epigenetically (by methylation, homologous RNAi inactivation, etc.). There is nothing out of the ordinary in this speculation, since most DNA in the human genome is likely being transcribed anyway (Wong et al. 2001). It is also plausible that some degraded transcripts of pseudogenes-to-be and pseudogenes are rejected at pre-translation checkpoints (like the presence of 5' cap, nonsense-mediated decay, etc.) to minimize the load of dysfunctional proteins from abundant junk duplicates.

## Methods

The sequence data for protein-coding genes were retrieved from NCBI web site for human (*Homo sapiens*), rat (*Rattus norvegicus*), mouse (*Mus musculus*), fish (*Fugu rubripes*), nematode (*Caenorhabditis elegans*), fly (*Drosophila melanogaster*), plant (*Arabidopsis thaliana*) and yeast (*Saccharomyces cerevisiae*).

Gene duplicates were identified at the amino acid level by matching each gene with all others in a proteome and by clustering similar genes using the "blastclust" program available at NCBI. A gene of inquiry was considered as belonging in a given cluster of duplicates if its alignment with at least one another gene already included in the cluster spans no less than 60% of its length and has more than 50% amino acid identity. These homology criteria numbers represent not too divergent gene duplicates and a sufficiently long similarity region in order to make the analysis of nucleotide alignments sensible. To avoid saturation effects we limit considerations to duplicates younger than 0.6 substitutions per neutral site. We experimented with a few other criteria and found that the results are sufficiently invariant.

After clustering duplicates, we formed the sets of gene pairs for analyses. This is simple for clusters consisting of only two genes. However, for clusters of larger sizes, there is a problem of over-representation, since different pairs from one cluster may represent the same duplication event, thus not being independent. We experimented with different cluster sizes and found that the trends we observed do not strongly depend on the size of clusters. In presented data, we rather arbitrary chose clusters of size less than 6 to avoid bias from few large multigene families. These clusters cover about 80% of gene duplicates so that we covered the majority of duplication events.

To evaluate the selection pressure we used two techniques: one is traditional counting of aminoacid replacement substitutions (R) versus silent substitutions (S). The ratio of R/S reflects selective constraints on protein. At first two corresponding amino acid sequences were aligned by the "BestFit" utility from Wisconsin Package(TM). It uses a local homology algorithm described in (Smith and Waterman 1981). Based on that amino acid alignment, we created the nucleotide sequences alignment by our Perl script. This nucleotide alignment was then fed to the "Diverge" utility from the Wisconsin Package, which estimated the pair wise number of synonymous and non-synonymous substitutions per site based on the method described in (Li et al. 1985).

Another method exploited the fact that $1^{st}$, $2^{nd}$ and $3^{rd}$ codon positions in average are under different constraints, $2^{nd}$ position being the most conservative, $3^{rd}$ largely synonymous and $1^{st}$ intermediate. In this method, we aligned nucleotide sequences of duplicated genes by "bl2seq" program (from "blast" programs family available at NCBI), dropping the unalignable regions. These methods gave similar trends in the moderately diverged duplicates region but the advantages of the later are that we can directly compare the rates in all three codon positions and the angle between two streams of pseudogenes-to-be and functional duplicates is larger (Fig. 1-3).

Data for the rates of neutral substitutions for Fig. 6 was taken from (Yi et al. 2002; Stein et al. 2003; Tamura et al. 2004).

## Acknowledgements

We thank G. Holmquist for stimulating discussions and comments on the manuscript. Our special thank is to S. Bates for reading the manuscript.

**Table 1 - Duplications parameters for different species**

Parameters of fits by Eqs. (1,2) for different species.

|  | $k$ | $k_{23}$ | $N_p / sps_3$ | $N_f / sps_3$ | $\theta = \dfrac{N_p}{N_f}$ |
|---|---|---|---|---|---|
| Human | 19 | 0.23 | 22000 | 1600 | 14 |
| Mouse | 17 | 0.22 | 21000 | 2000 | 11 |
| Rat | 17 | 0.22 | 16500 | 1500 | 11 |
| Fugu | 27 | 0.21 | 13000 | 1900 | 6.8 |
| Nematode | 28 | 0.25 | 3600 | 1700 | 2.1 |
| Fly | 44 | 0.24 | 4000 | 520 | 7.7 |
| Yeast | 45 | 0.18 | 1500 | 290 | 5.2 |
| Weed | 29 | 0.21 | 5300 | 4400 | 1.2 |

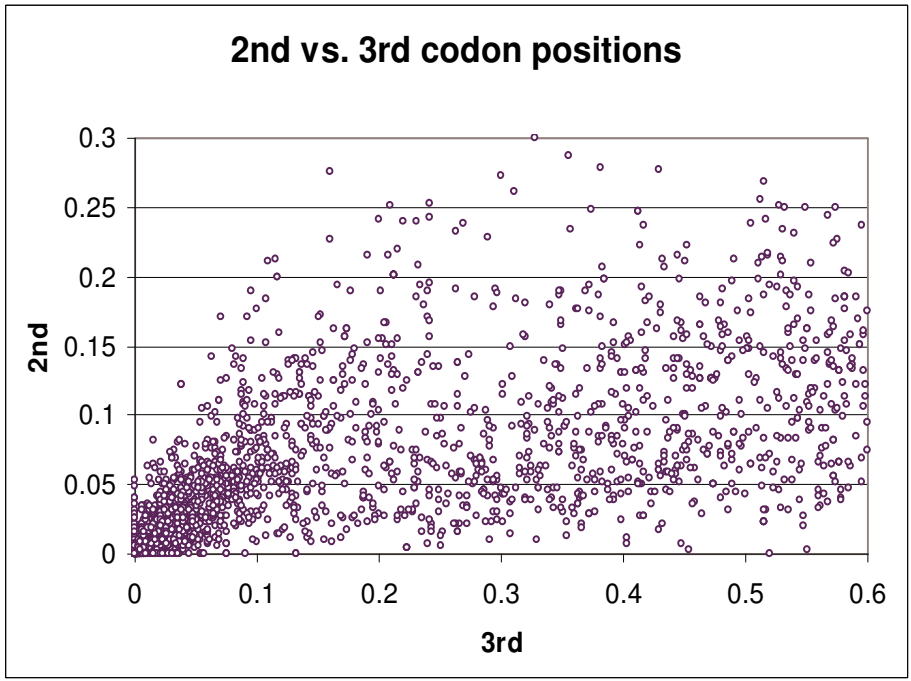 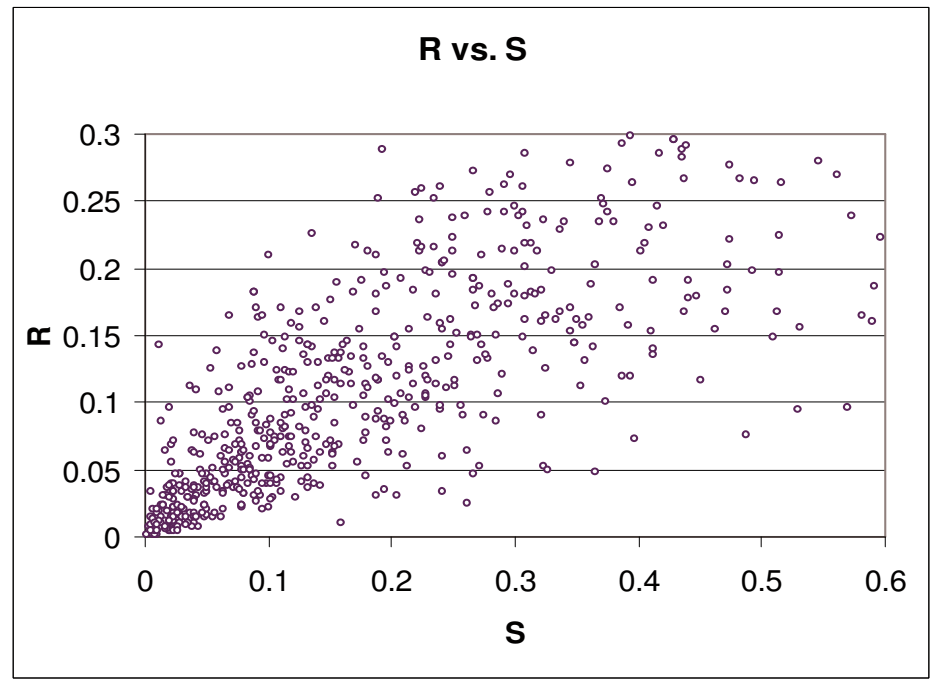

**Figure 1 - Divergence of human duplicates**
Each point represents a divergence of a pair of duplicates measured with corresponding metrics.

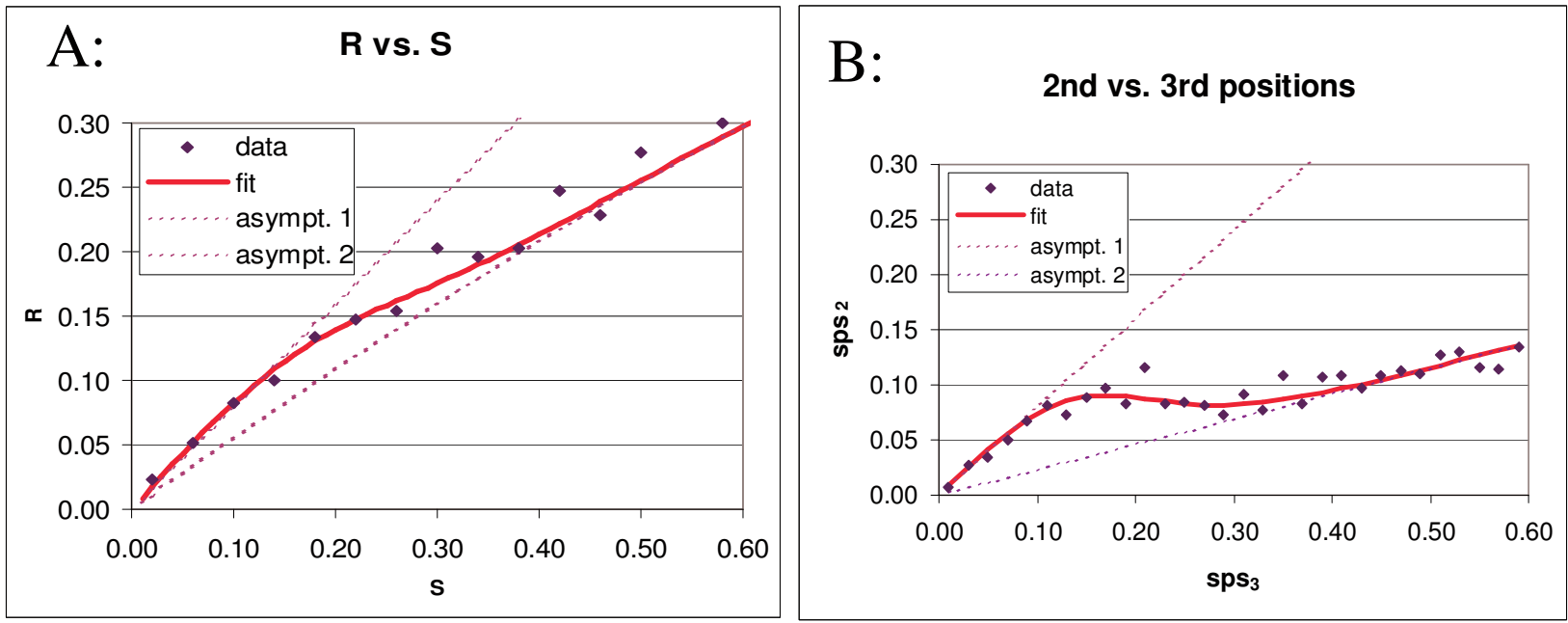

**Figure 2 - Averaged selection pressure on human duplicates represented by 2nd vs. 3rd codon positions and R vs. S**

Each data point represents an averaged number of corresponding *sps* with X-axis bin size 0.02. Red solid lines are fits by the models discussed in text. Dashed lines are asymptotes of substitutions rates for neutral and purifying sequence evolution (young and old duplicates correspondingly).

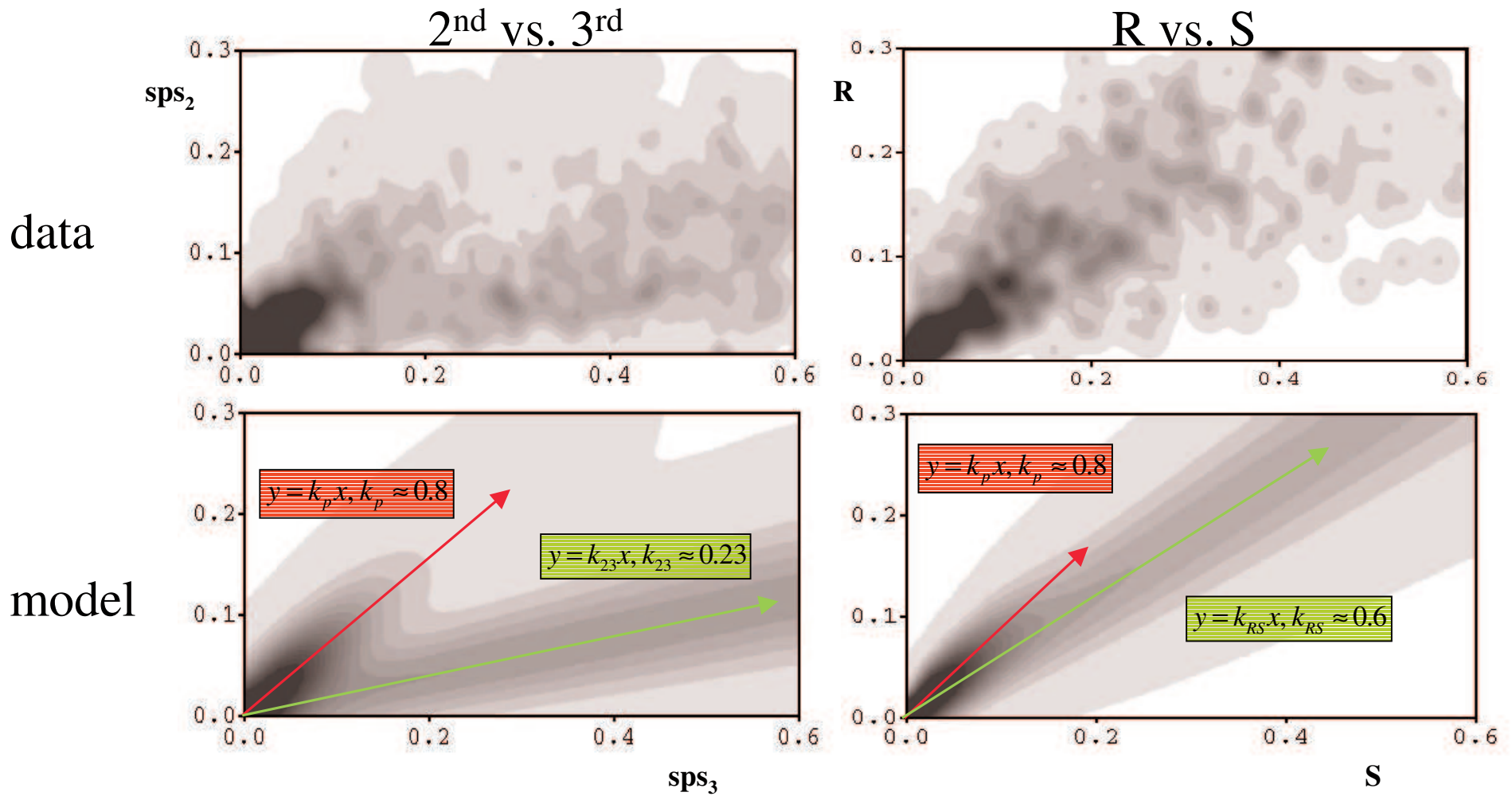

**Figure 3 - Density plots of human duplicates and two-streams model**

The plots represent the density of duplicates in divergence space analogous to Fig. 1. Bottom plots – the model described by Eqs. 1, 2 with variance (corresponds to streams "width") modelled by binomial distribution. Top – human duplicates data from Fig. 1 smeared by convolution with gauss distribution. Decaying stream along $y = k_p x$, $k_p \approx 0.8$ represents pseudogenes-to-be being wiped out from genome by mutagenesis. A second steady stream for 2nd vs. 3rd codon positions $y = k_{23} x$, $k_{23} \approx 0.23$ represents functional genes under purifying constraints. Corresponding slope for R vs. S metric is 0.6.

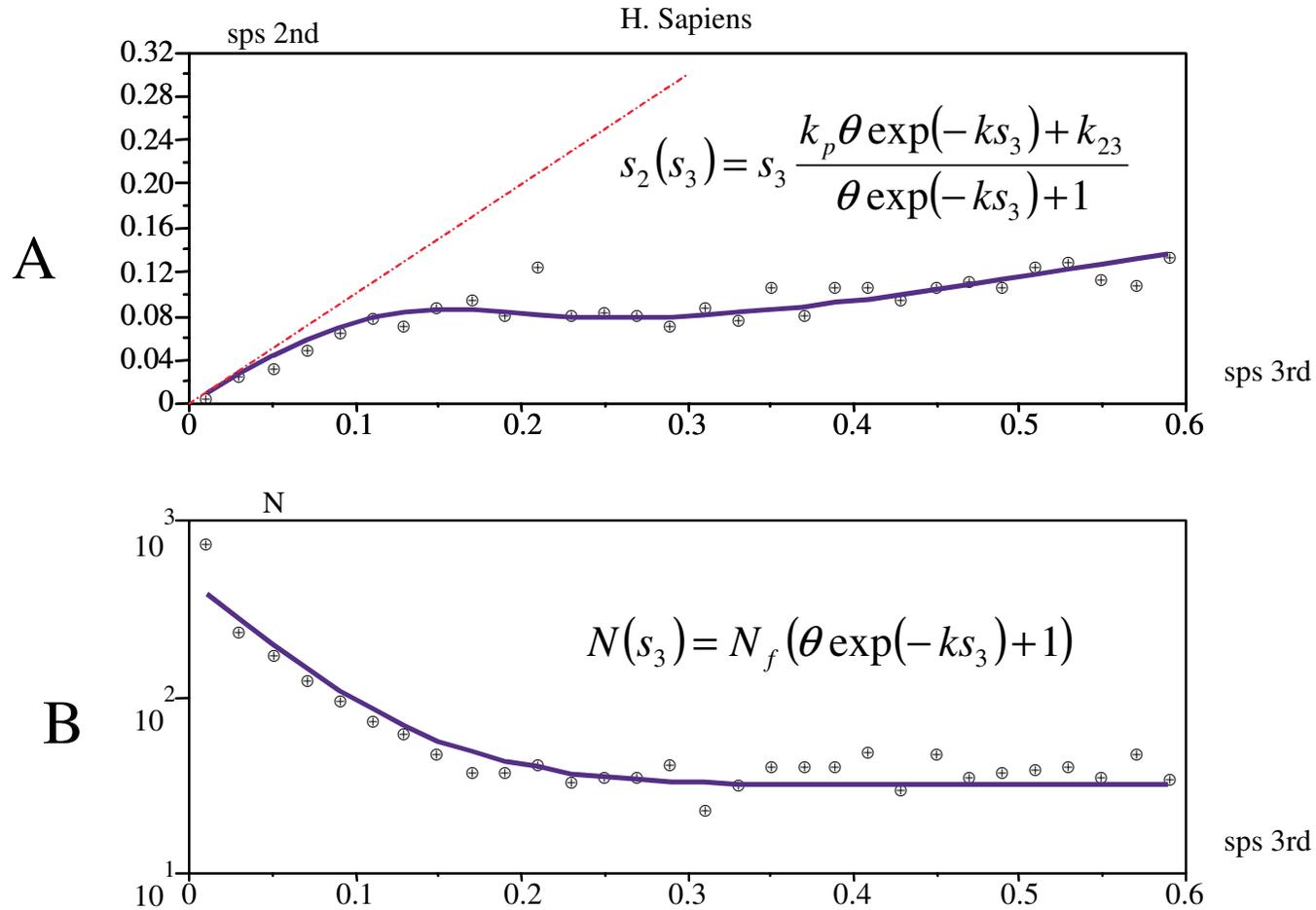

**Figure 4 - Model fit for human duplicates**

The fits of Eqs. 1 (B) and 2 (A) for duplicates of *Homo sapiens*. A plot: averaged sps number of 2[nd] vs. 3[rd] codon positions; reflecting selection pressure; lower - duplicates number vs. sps in 3[rd] codon positions. Averaging bin size is 0.02. Fit parameters are: $k=19$, $k_{23}=0.23$, $N_f / sps_3 = 1600$, $\theta = 14$. Red line shows *y=x* slope.

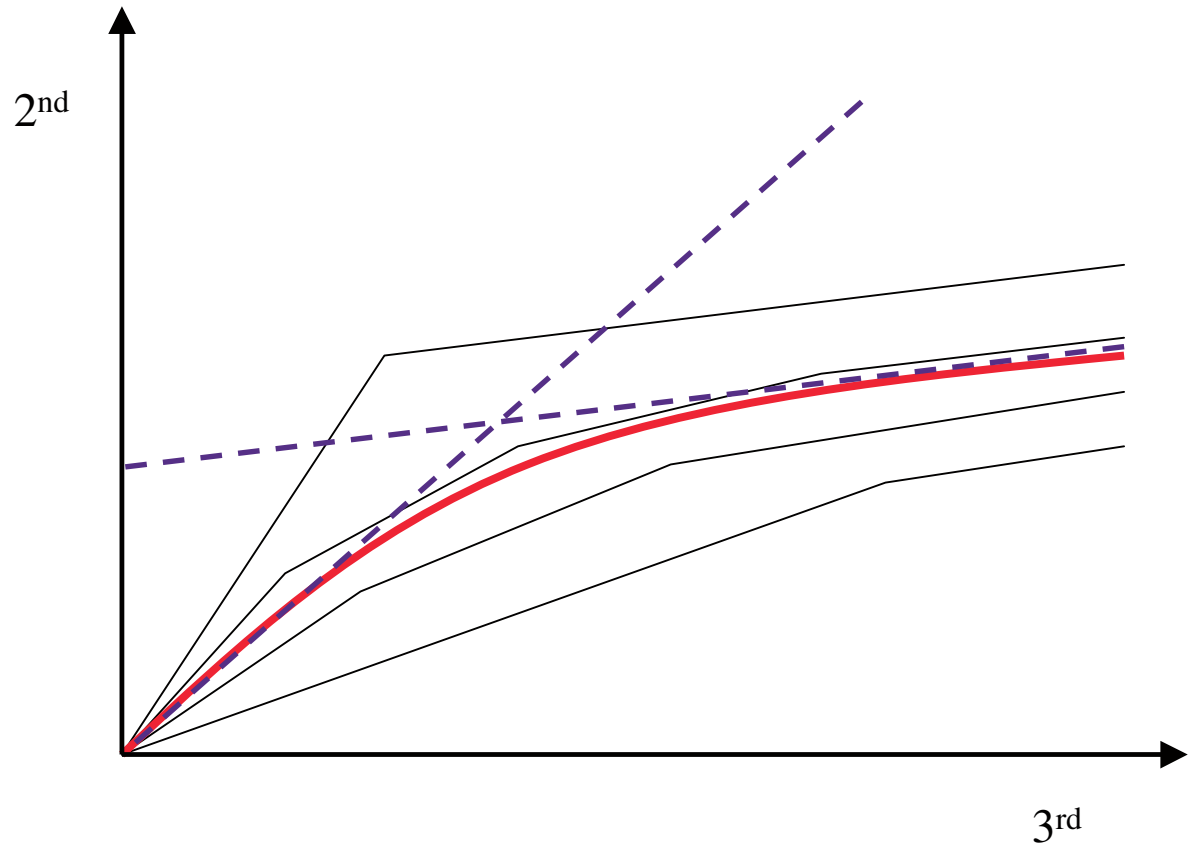

**Figure 5 - Expected average trend for the gradual change of constraints model**

If we assume that the majority of duplicates follow the gradual paths (black trajectories) from the weak constraints to the strong ones, we could expect the only possible average trend as shown by the red curve – positive slope and negative curvature. And the asymptote (dashed line) for old duplicates could not point to coordinate origin (compare to Fig. 2B). Hence the gradual changes assumption contradicts with the pattern with a "saddle region", we observe in all large genomes studied (Fig. 2B).

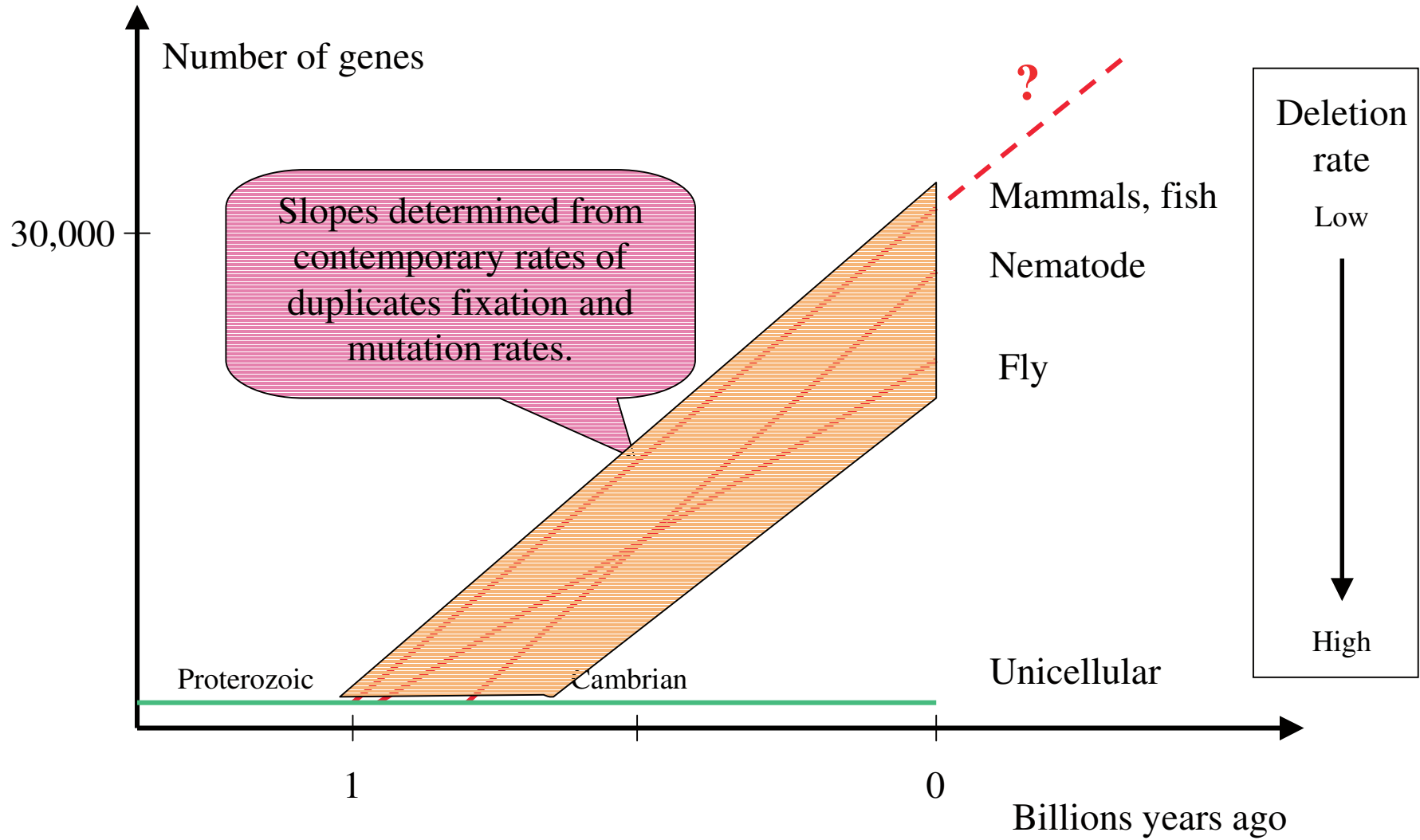

**Figure 6 - Clock property of the rates of functional duplicates fixation.**

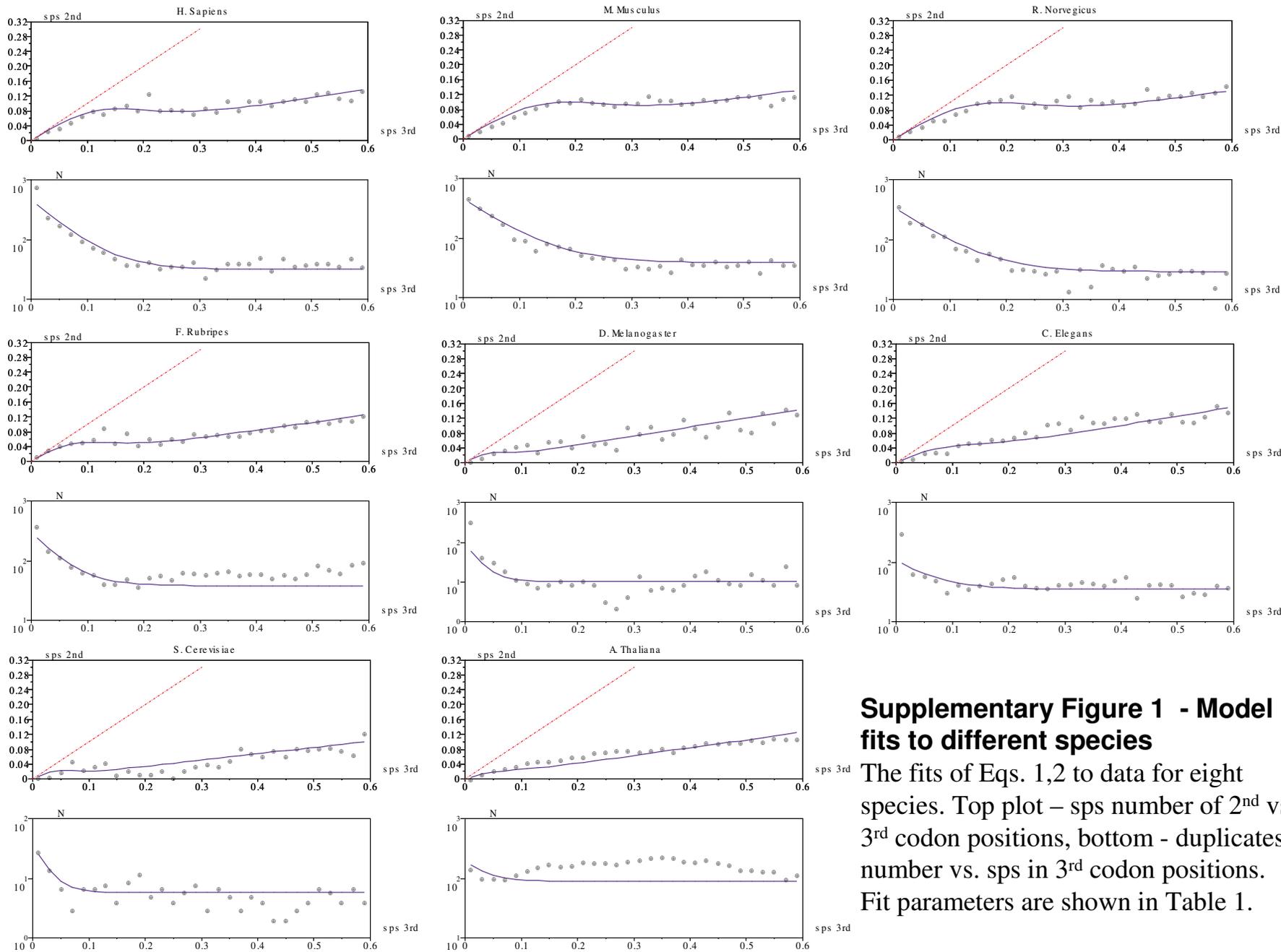

**Supplementary Figure 1 - Model fits to different species**

The fits of Eqs. 1,2 to data for eight species. Top plot – sps number of $2^{nd}$ vs. $3^{rd}$ codon positions, bottom - duplicates number vs. sps in $3^{rd}$ codon positions. Fit parameters are shown in Table 1.